\documentclass[pra,longbibliography,twocolumn,showpacs,nofootinbib,superscriptaddress,notitlepage]{revtex4-2}

\pdfoutput=1 

\usepackage[ruled,vlined]{algorithm2e}

\usepackage{dsfont}
\usepackage{amsmath}
\usepackage{amssymb,bm}
\usepackage{amsthm}
\usepackage{mathtools}
\usepackage{url}

\usepackage{array, makecell}
\usepackage{boldline}
\usepackage{enumitem}

\usepackage{color,dsfont}
\usepackage{graphicx}
\usepackage{ragged2e}
\usepackage[colorlinks=true, hyperindex, breaklinks, linkcolor=blue, urlcolor=blue, citecolor=blue]{hyperref} 
\usepackage[normalem]{ulem}
\usepackage[capitalise]{cleveref}
\usepackage{mathrsfs}
\usepackage[caption=false]{subfig}
\usepackage{mathtools}
\usepackage{float}
\usepackage{verbatim}
\usepackage{latexsym}
\usepackage{amsmath}
\usepackage{amssymb}
\usepackage{setspace}
\usepackage{amsfonts}
\usepackage{stmaryrd}
\usepackage{xcolor}
\usepackage{enumitem}
\usepackage{lipsum}

\newcommand{\ket}[1]{| {#1} \rangle }

\usepackage{multirow}
\usepackage{booktabs}

\begin{document}
\title{A circuit-level protocol and analysis for twist-based lattice surgery}
\author{Christopher Chamberland}\email{cchmber@amazon.com}
\affiliation{AWS Center for Quantum Computing, Pasadena, CA 91125, USA}
\affiliation{IQIM, California Institute of Technology, Pasadena, CA 91125, USA}
\author{Earl T. Campbell}\email{earltcampbell@gmail.com}
\affiliation{AWS Center for Quantum Computing, Cambridge, UK}
\begin{abstract}
Lattice surgery is a measurement-based technique for performing fault-tolerant quantum computation in two dimensions.  When using the surface code, the most general lattice surgery operations require lattice irregularities called twist defects.  However, implementing twist-based lattice surgery may require additional resources, such as extra device connectivity, and could lower the threshold and overall performance for the surface code.  Here we provide an explicit twist-based lattice surgery protocol and its requisite connectivity layout. We also provide new stabilizer measurement circuits for measuring twist defects which are compatible with our chosen gate scheduling. We undertake the first circuit-level error correction simulations during twist-based lattice surgery using a biased depolarizing noise model. Our results indicate a slight decrease in the threshold for timelike logical failures compared to lattice surgery protocols with no twist defects in the bulk.  However, comfortably below threshold (i.e. with CNOT infidelities below $5 \times 10^{-3}$), the performance degradation is mild and in fact preferable over proposed alternative twist-free schemes. Lastly, we provide an efficient scheme for measuring $Y$ operators along boundaries of surface codes which bypasses certain steps that were required in previous schemes. 
\end{abstract}
\maketitle

\section{Introduction}
\label{sec:Intro}

The surface code is a quantum error correcting code with a high threshold and which can be realised with two-dimensional hardware~\cite{dennis02,fowler2012surface}.  Other codes have been observed to have comparable thresholds for toy noise models \cite{jochym2014using,Yoder2016,ChamberlandPRL,chamberland2017overhead,ChamberlandColorCode,beverland2021cost}, but when considering realistic circuit-level noise their performance (relative to the surface code) drops appreciably.  The surface code's high threshold can be partially attributed to the simplicity of its stabilizers and the circuits used to measure them. Realising the surface code requires measurements of pure $X$-type and $Z$-type stabilizers involving no more than four qubits. These measurements can be performed using a depth-four sequence of two-qubit gates; the minimum we can hope for.  Furthermore, choosing the right gate schedule can make the stabilizer measurement circuits immune to ``hook" errors that effectively reduce the code distance. The moral here is that careful, circuit-level analysis of the microscopic details is crucial for a full understanding of the best route to fault-tolerant quantum computing.

Logical operations in two-dimensional surface code architectures can be performed by code deformation where the code stabilizers change over time, such as through braiding~\cite{Raussendorf06,fowler2012surface} or lattice surgery~\cite{horsman2012surface,litinski2018lattice,Vuillot_2019}, with the latter being more resource efficient~\cite{fowler2018low,litinski2019game}. The basic primitive in lattice surgery is a logical multi-qubit Pauli measurement, which when combined with logical ancilla states can be used to fault-tolerantly perform logical Clifford operations in addition to logical non-Clifford gates by teleportation. When measuring logical Paulis composed of $X$ and $Z$ operators, the code is deformed with several surface code patches merged into a large patch, though locally the code stabilizers are unchanged. However, more generally, measurements involving Pauli $Y$ operators are needed. Such measurements can be realised with twist-based lattice surgery~\cite{litinski2018lattice}, though this requires some weight-five measurements in addition to long range multi-qubit gates. The proposed twist-based approaches were presented abstractly, ignoring circuit-level implementation details and any related performance impacts. 

\begin{figure}[b!]
    \centering
    \includegraphics[width=0.8\columnwidth]{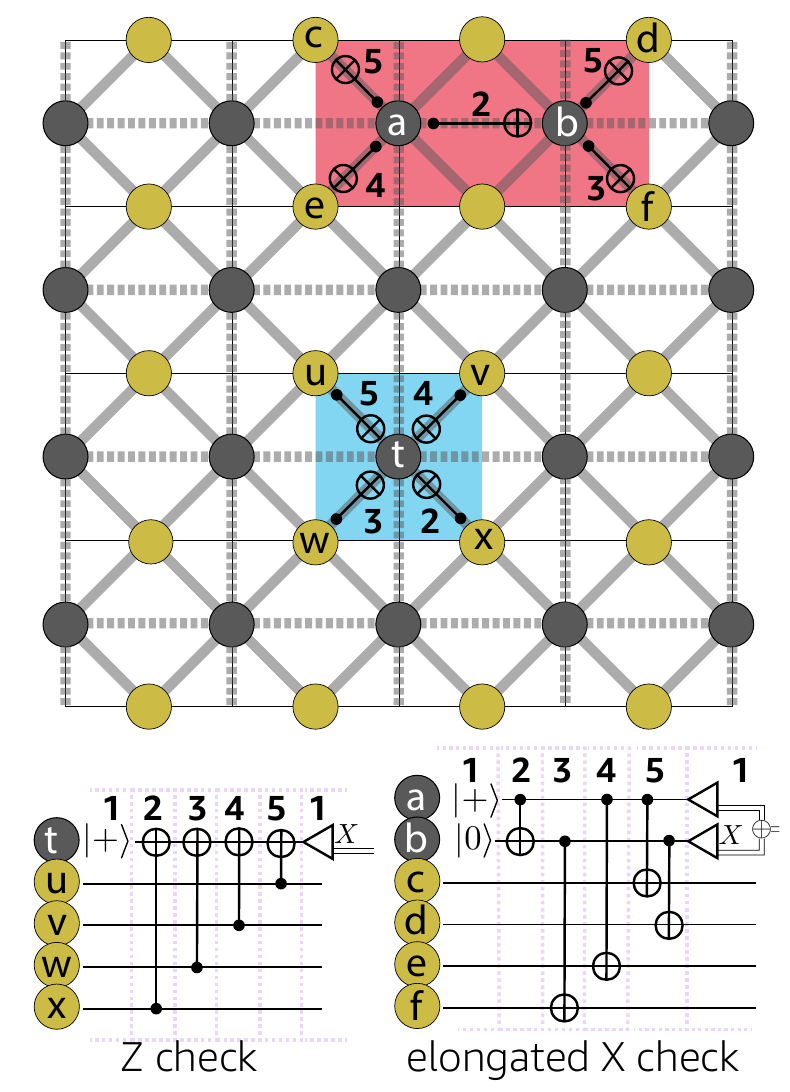}
    \caption{Connectivity diagram between data qubits (yellow) and ancilla qubits (grey) with two example gate schedules.  Thick, solid grey lines show the connectivity needed to perform a standard weight-four check as shown here for a blue $Z^{\otimes 4}$ stabilizer. To perform a check along an elongated rectangle (such as the red check shown in the figure) additional connectivity can be used which we illustrate as dashed grey lines. For the elongated $X^{\otimes 4}$ check, the two grey ancilla qubits are prepared in a GHZ state which is used to perform the desired stabilizer measurement. Examples of gate schedulings and circuits used to measure stabilizers are shown (numbers indicate the time step for each two-qubit gate) and further circuit details are given in \cref{appendix:StabMeasTwist}. The first and last time steps are used for state preparation and measurement and so are both labeled as time step 1. In subsequent figures, we will sometimes just provide numerical labels when the type of gate is clear from the context. }
    \label{fig:Connectivity}
\end{figure}

Our work gives the first proposal for a circuit-level implementation of twist-based lattice surgery and numerically benchmarks the performance.  In twist-based lattice surgery, we need to measure some longer range stabilizers that we call elongated rectangles and twist defects. To perform these measurements without increasing the circuit-depth (which would significantly degrade performance), we assume the hardware connectivity illustrated in \cref{fig:Connectivity}. We say two qubits are connected when a two-qubit gate can be performed between these qubits.  Standard surface code implementations have each qubit connected to four other qubits. In this work, we assume a minimal extension where some qubits are connected to eight other qubits. The cat-qubit based architecture of Ref.~\cite{chamberland2020building} naturally provides this connectivity, and it could be produced in other architectures, though at the risk of additional cross-talk. Yoder and Kim proposed a twist-based scheme with lower weight measurements~\cite{YoderKim17}, but they required a substantial change to the underlying surface code~\cite{YoderKim17} and did not use a homogeneous (translationally invariant) hardware. We leave open the question of whether twist-based lattice surgery can be realised using less connectivity within a homogeneous two-dimensional architecture.

To benchmark twist-based lattice surgery, we perform Monte Carlo simulations with a minimum-weight perfect matching (MWPM) decoding protocol (following the method of Ref.~\cite{CC21}) of $Y \otimes Y$ and $Z \otimes Z$ measurements.  We perform our simulations for biased noise and rectangular surface codes since this enables us to perform larger simulations and these noise models are of recent interest~\cite{TuckettBiasedNoise,XZZXcodes,BravyiBiased,PuriXZZX,HiggottSubsystem,chamberland2020building}.  We observe that the measurement of a logical $Y \otimes Y$ operator has a higher logical failure rate relative to a comparable $Z \otimes Z$ measurement of the same area. We attribute such differences to the presence of twist defects and elongated checks.  The logical failure probability of these measurements is exponentially suppressed in the number of stabilizer measurement rounds used.  Therefore, we can ask how many more rounds (and therefore time cost) is needed for a $Y \otimes Y$ measurement to achieve the same logical fidelity as a $Z \otimes Z$ measurement. We find, well below threshold ($p_{\text{CNOT}} \leq 0.1 \%$), the use of twists incurs only a small multiplicative time cost (less than $1.2\times$ for one of the studied noise models when $p_{\text{CNOT}} = 10^{-3}$) since the twists are a small fraction of all possible fault locations.  However, the twist defects are the weakest point and thus fail with a higher probability compared to other fault locations. Consequently, we find these multiplicative time costs increase as we approach the threshold. 

In previous work~\cite{CC21}, we proposed an alternative twist-free lattice surgery scheme that replicated the computational power of Pauli $Y$ measurements without the need for twist defects.  However, this alternative strategy came with its own $\sim 2\times$ time cost. Therefore, a key conclusion of this work is that twist-based approaches outperform twist-free alternatives when operating comfortably below threshold, and under our connectivity assumptions along with the biased noise model used for the simulations. Twist-free lattice surgery may still prove useful in connectivity limited hardware or very close to threshold. Hardware afflicted by a different noise model than considered in this work might also benefit from twist-free lattice surgery.

\begin{figure*}
    \centering
    \includegraphics[width=\textwidth]{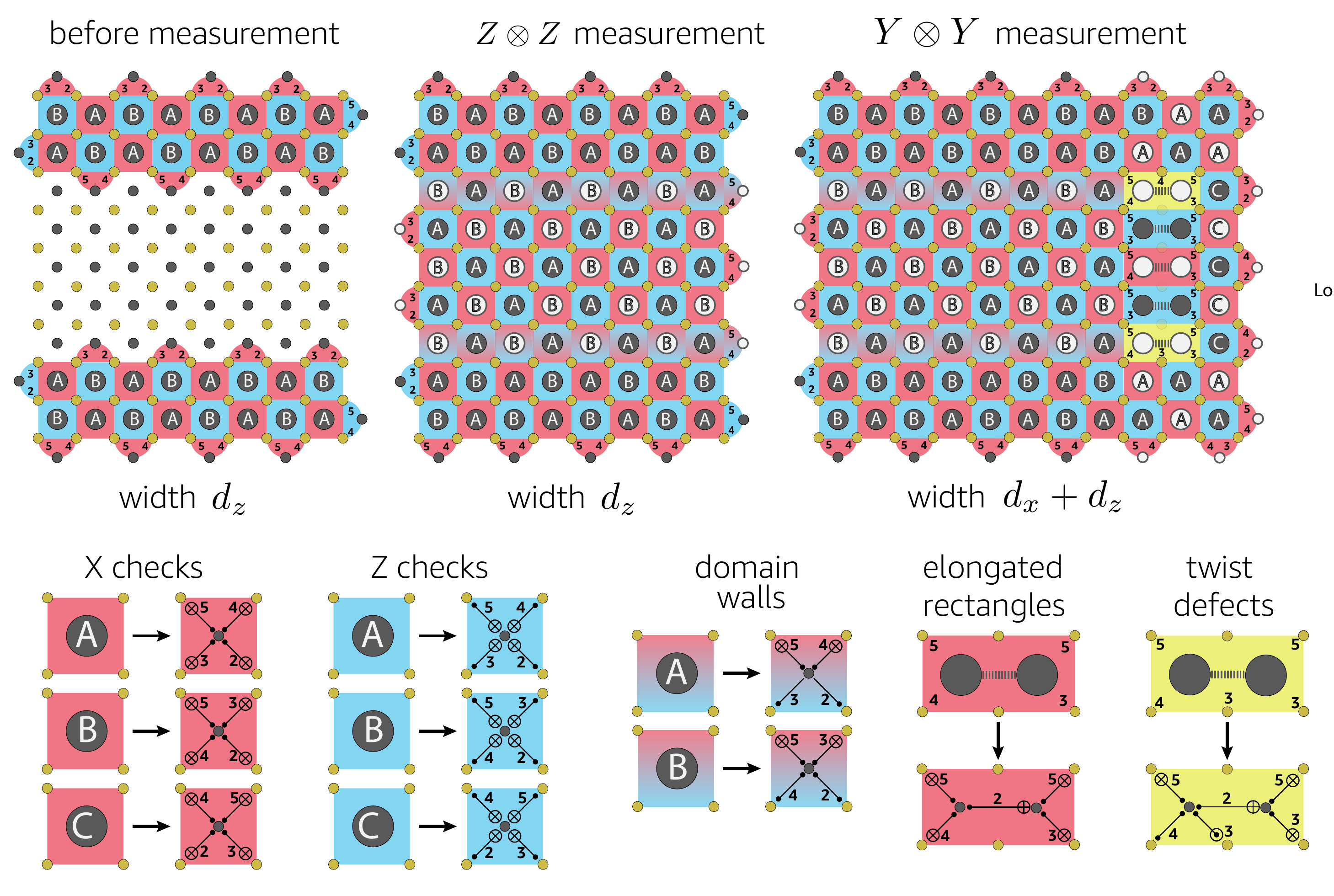}
    \caption{Illustration of a $Z \otimes Z$ and $Y \otimes Y$ measurement performed by lattice surgery. The left most figure shows two $d_x = 3$ and $d_z = 9$ logical patches prior to the measurements. The qubits (yellow and grey vertices) in between the two patches are part of the routing space. The middle figure shows the measurement of $Z \otimes Z$ by gauge fixing, i.e. measuring the appropriate stabilizers in the routing space such that a $Z \otimes Z$ measurement is performed along the logical $Z$ boundaries of the surface code patches. Red plaquettes correspond to $X$-type stabilizers, blue plaquettes to $Z$-type stabilizers. Gradient colour plaquettes correspond to domain walls that measure mixed $X$ and $Z$-type stabilizers. The labels A, B and C describe particular gate schedulings. Stabilizers with white vertices encode the parity of the logical $Z \otimes Z$ measurement. The rightmost figure illustrates a logical $ Y\otimes Y$ measurement implemented by lattice surgery. As explained in \cref{sec:FaultToleranceGauge}, the $Y \otimes Y$ measurement can be done directly without first extending the logical patches so that logical $Y$ operators can be expressed along their respective horizontal boundaries. When performing the gauge fixing step, weight-five stabilizers illustrated by yellow plaquettes (referred to as twist defects) must be measured, in addition to elongated $X$ and $Z$-type stabilizers. The measurements of both twist defects and elongated stabilizers require the use two ancilla qubits prepared in a GHZ state with the assumed hardware connectivity constraints described in \cref{sec:Intro}. Lastly, we remark that in addition to the A and B gate schedules, the $Y \otimes Y$ also requires the use of the C schedule.  }
    \label{fig:JustYY}
\end{figure*}

\section{Lattice surgery with twist defects in the bulk}
\label{section:CYYsec}

In this section, we describe our protocol for performing twist-based lattice surgery. In \cref{subsec:GateScheduling}, we provide a gate scheduling that can be used to measure an arbitrary number of Pauli $Y$ operators using twist defects in the bulk. Fault-tolerant circuits for measuring stabilizers at twist defects are provided in \cref{appendix:StabMeasTwist}. In \cref{subsec:PerformanceCompare}, we will then compare the performance of a lattice surgery protocol for measuring $P = Y \otimes Y$ to a $P = Z \otimes Z$ protocol that does not require twist defects in the bulk. The Pauli $Z \otimes Z$ is chosen for comparison since the measurement of both $Y \otimes Y$ and $Z \otimes Z$ require domain walls between the logical patches and the routing space. In what follows, the surface code encoding a logical qubit will be referred to as a logical patch. The extra space required to perform lattice surgery will be referred to as the routing space.

\subsection{Gate scheduling for twist-based lattice surgery}
\label{subsec:GateScheduling}

Recall that in a lattice surgery protocol, mutli-qubit Pauli measurements between logical patches are performed by measuring a set of operators in the routing space between the logical patches. Such measurements (which correspond to gauge fixing \cite{Vuillot_2019}) merge the logical patches into one large surface code patch, and a subset of the stabilizers of the merged patch encode the parity of the logical multi-qubit Pauli measurement. A simple example for a $Z \otimes Z$ and $Y \otimes Y$ measurement is provided in \cref{fig:JustYY}. 

\begin{figure}
    \centering
    \includegraphics[width=0.95\columnwidth]{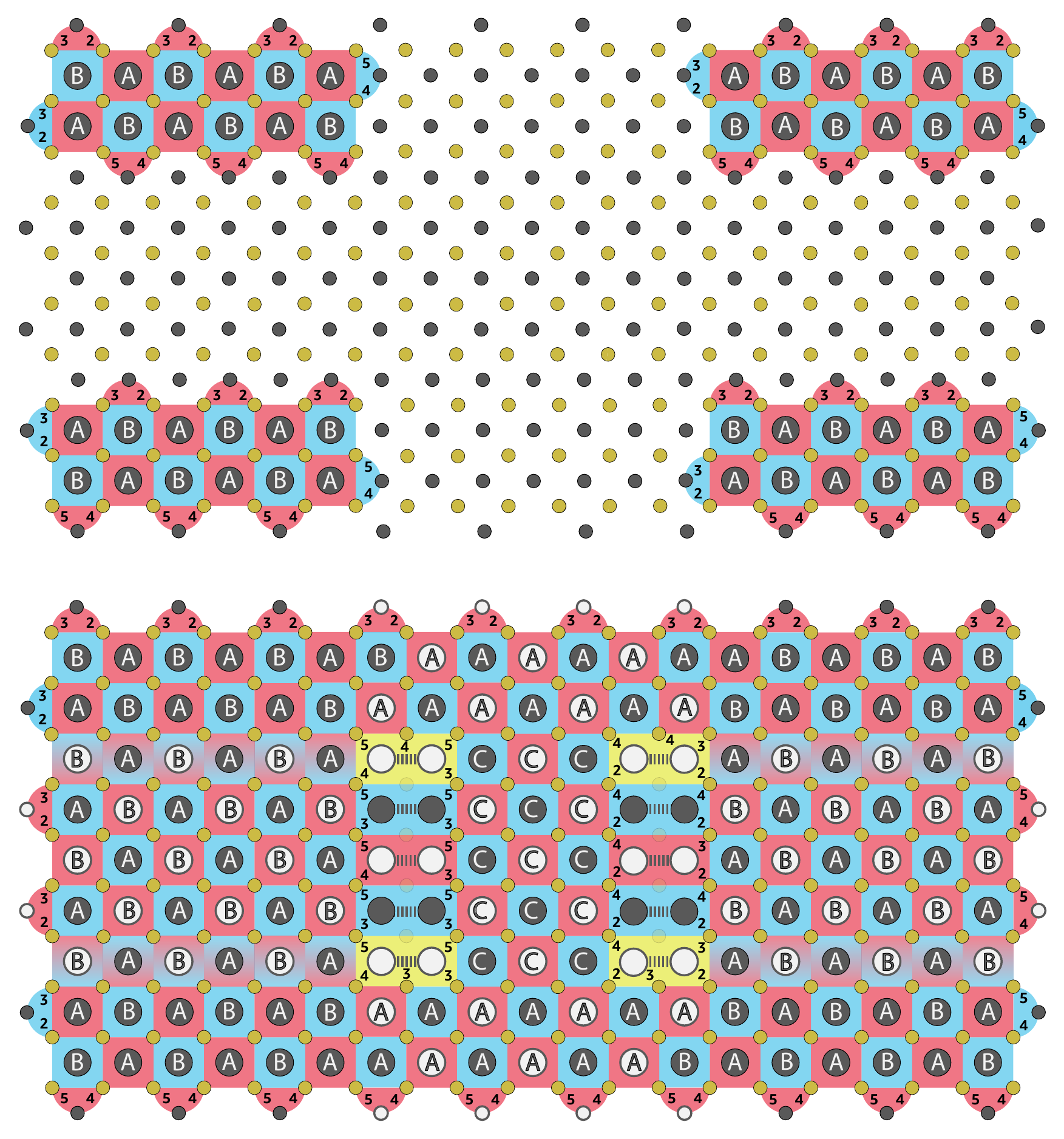}
    \caption{Lattice and valid gate scheduling for measuring the multi-qubit Pauli operator $P = Y \otimes Y \otimes Y \otimes Y$ via lattice surgery. The A, B and C gate schedules are described in \cref{fig:JustYY}. The chosen gate scheduling imposes the need for different ancilla states used to measure twist defects on the left strip compared to twist defects on the right strip. More details are provided in \cref{appendix:StabMeasTwist}.}
    \label{fig:GaintYY}
\end{figure}

Let $P = P_1 \otimes P_2 \otimes \cdots \otimes P_k$ be a multi-qubit Pauli operator that is to be measured using a lattice surgery protocol. If $P_j \in \{X,Z\}$ for all $j \in \{1, \cdots , k\}$, then the standard surface code gate scheduling can be used to measure all the stabilizers during the lattice surgery protocol \cite{TomitaSvore}. For instance, see the scheduling used for a logical $Z \otimes Z$ measurement performed by lattice surgery in \cref{fig:JustYY}. With such a scheduling, all two-qubit gates can be applied in four time steps thus minimizing the number of idling qubit locations. One of the important features of the standard surface code gate scheduling circuits is that a weight-two error arising from a single fault will be perpendicular to the relevant logical operator thus preserving the surface code distance (i.e. the code will be guaranteed to correct any error arising from at most $(d-1)/2$ faults). By relevant logical operator, we mean that if the weight-two data-qubit error arising from a single fault is of the form $E_{X} = X_{q_1} \otimes X_{q_2}$ and the minimum-weight representatives of $X_L$ form vertical strings, then $E_{X}$ will have support along horizontal strings. Similarly, $E_{Z}$ will have support along vertical strings since the minimum-weight representatives of $Z_L$ form horizontal strings.

Now suppose there exists at least one $j \in \{1, \cdots, k \}$ such that $P_j = Y$. In Fig.2d of Ref~\cite{litinski2019game}, Litinski proposed measuring the $Y$ operators as follows. First, logical patches are extended using qubits in the routing space such that the logical $Y_L$ operator of the logical patch can be expressed along a horizontal boundary. The logical $Y$ operator is then given by 
\begin{equation}
    Y = Z_{q_1} \otimes Z_{q_2} \otimes \cdots \otimes X_{q_{d_z-1}} \otimes Y_{q_{d_z}}  \otimes X_{q_{d_z + 1}} \otimes \cdots \otimes X_{q_{d_z + d_x}}
\end{equation} 
where $d_x$ and $d_z$ are the minimum weights of the logical $X$ and $Z$ operators of the logical patch. The $Y$ operator can then be measured via gauge fixing by performing $Z$ and $X$-type stabilizer measurements between the extended logical patch and qubits in the routing space. The gauge-fixing step requires the measurement of mixed $X \otimes X \otimes Z \otimes Z$ stabilizers which we refer to as domain walls. In addition, it requires the measurement of a weight-five operator operator containing a physical $Y$ term. Such weight-five operators are referred to as twist defects. An important feature of our scheme is that when measuring multi-qubit Pauli's involving $Y$ operators, such measurements can be done \textit{directly}, without first extending the logical patches.  Using an extension step doubles the implementation time and uses additional routing space, thus making our direct approach more efficient. Further, we show in \cref{sec:FaultToleranceGauge} that our direct approach for measuring Pauli $Y$ operators is fault-tolerant using the gauge fixing formalism of Ref.~\cite{Vuillot_2019}.

The presence of twist defects reduces the set of gate scheduling solutions that can be used for a twist-based lattice surgery protocol with minimal circuit depth. A valid gate scheduling for the measurement of $P = Y \otimes Y$ via lattice surgery is shown in \cref{fig:JustYY}. The chosen gate scheduling ensures that all two-qubit gates can be applied in four time steps. Three different gate schedules are used, which are labelled A, B and C. The C schedule is only used in the routing space to the right of the twist defects. Note that the A, B and C schedules must follow a set of rules for how they can be tilled on the lattice in order to avoid scheduling conflicts or invalid schedules for performing the stabilizer measurements\footnote{A valid schedule is one where all stabilizers are guaranteed to be measured as $+1$ in the absence of faults. Further, if an error anticommutes with a stabilizer $g$, then $g$ is guaranteed to be measured as $-1$.}. For instance, a stabilizer measured using an A schedule cannot be adjacent to a stabilizer measured using a C schedule in the horizontal direction. If they were, qubits shared by both stabilizers would interact with two-qubit gates in the same time steps. On the other hand, C and A schedules can be adjacent to each other in the vertical direction (see the legend at the bottom of \cref{fig:JustYY}). 

\begin{figure*}
    \centering
    \includegraphics[width=\textwidth]{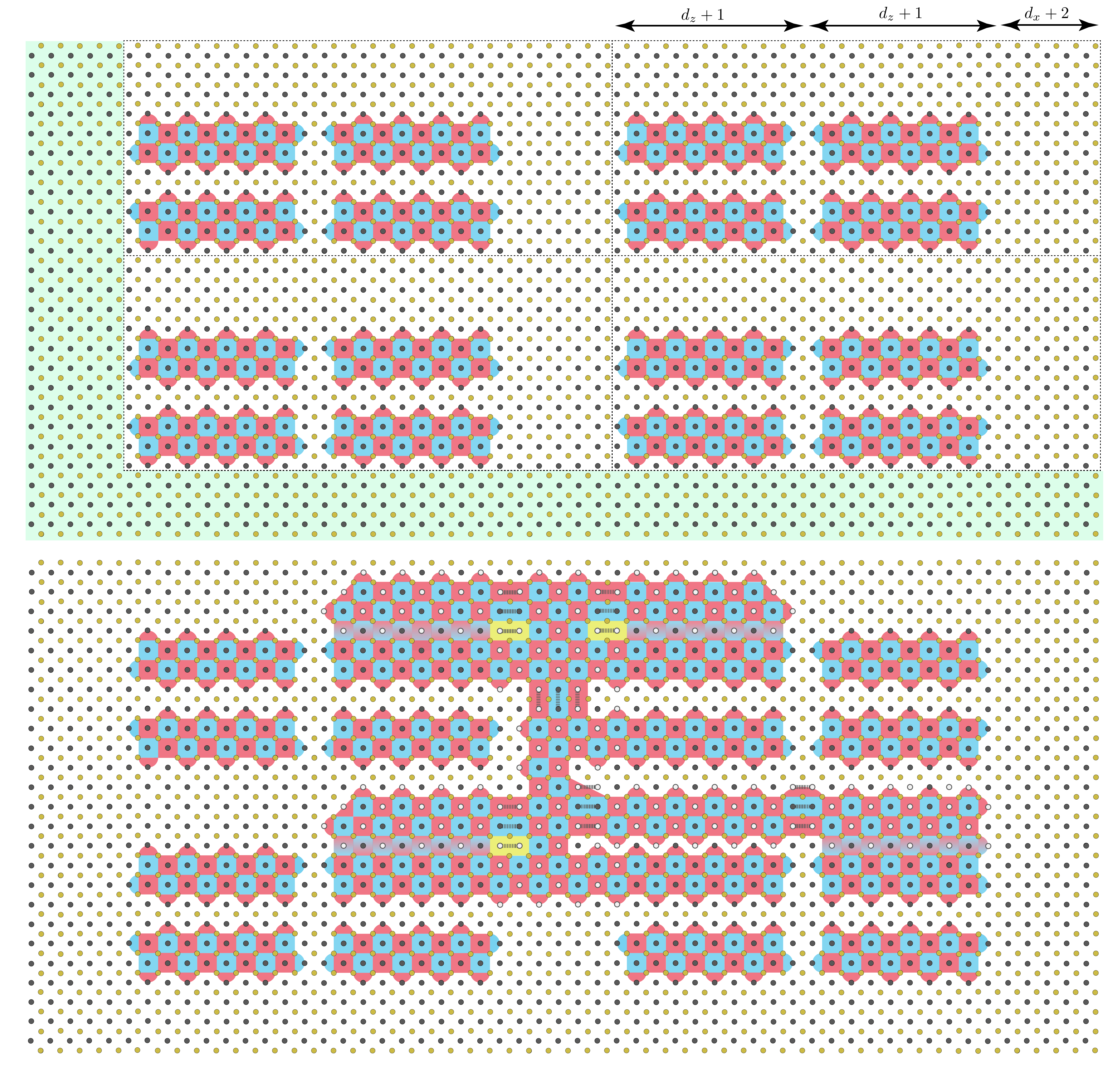}
    \caption{Example layout of 16 logical patches with routing overhead used for performing arbitrary multi-qubit Pauli measurements via lattice surgery. In the top figure, we illustrate the logical patches prior to performing the Pauli measurements. We illustrate 4 unit cells, with each unit cell (shown with a dashed line) containing 4 surface code patches and some routing space.  Some additional edge padding (shown in green) is also need for routing space.  In the bottom figure, we illustrate the merged patch that is obtained after measuring a logical $Y \otimes Y \otimes X \otimes Y \otimes X \otimes Z$ Pauli operator. The gate scheduling presented in \cref{fig:GaintYY} can be used for performing the three $Y$ measurements.  }
    \label{fig:IntroduceIdeas}
\end{figure*}

Unlike the standard surface code gate scheduling where all $X$-type stabilizers use the A schedule and $Z$-type stabilizers use the B schedule (or vice-versa), the gate scheduling for measuring $Y \otimes Y$ includes some regions where the A schedule is applied uniformly. Consequently, there will be single failures resulting in weight-two data qubit errors which are parallel to the logical $X_L$ or $Z_L$ operators of the merged patch. For a given quantum algorithm, the $d_x$ and $d_z$ distances of logical patches are chosen such that a single logical data qubit error is very unlikely during the course of the entire computation. It is important to note that the merged lattices of a twist-based protocol with distances $d_x'$ and $d_z'$ for measuring $Y$ operators will have $d_x' \geq d_x$ and $d_z' \geq d_z$ . In other words, the distances of the merged patches will be larger than the distances required for the logical patches prior to the merged. Therefore, even though our gate scheduling has the property that some weight-two errors arising from single failures will be parallel to $X_L$ or $Z_L$ logical operators, the increased effective distances during lattice surgery protocols will ensure that $d_x$ and $d_z$ distances of the un-merged logical patches remain unchanged when performing twist-based lattice surgery, which we observe numerically. There are, however, some subtleties when performing twist-based lattice surgery that affect the minimum required value of $d_x$ (see \cref{app:Subtleties} for details).

We now provide a generalization of the gate scheduling shown in \cref{fig:JustYY} used for twist-based lattice surgery. In \cref{fig:GaintYY}, we illustrate the gate scheduling required to measure the logical Pauli operator $P = Y \otimes Y \otimes Y \otimes Y$. The inclusion of a second pair of $Y$ operators has the feature that all stabilizers of the merged patch in the routing space between the four twist defects are measured using the C schedule. As was the case in \cref{fig:JustYY}, the presence of twist defects requires the measurement of elongated stabilizers along vertical strips located between the twist defects in the routing space. In order to respect the nearest neighbor connectivity constraints imposed by many quantum hardware architectures, additional ancilla qubits need to be used to measure the elongated stabilizers and weight-five operators. However, an important feature of the circuit in \cref{fig:GaintYY} is that the gate scheduling imposes the need for different syndrome measurement circuits for the twist defects on the left strip compared to those on the right strip. More details are provided in \cref{appendix:StabMeasTwist} with the circuits shown in \cref{fig:StabMeasCircuitsTwist}.

We remark that the gate scheduling shown in \cref{fig:GaintYY} can be extended to measure arbitrary multi-qubit Pauli operators such as the one shown in \cref{fig:IntroduceIdeas}. In particular, \cref{fig:IntroduceIdeas} provides the minimal routing space requirements to perform arbitrary multi-qubit Pauli measurements for logical patches placed on a two-dimensional grid with qubits afflicted by a biased noise model.  We assess the routing overhead by considering an individual \textit{unit cell} of 4 logical patches shown in \cref{fig:IntroduceIdeas}.  Each unit cell uses $8 d_x d_z$ physical qubits (which includes yellow and grey vertices) for the four logical patches and contains a total of $2(3 d_x +2)(2 d_z + d_x +4)$ physical qubits. The factor $3 d_x +2$ is the height of the unit cell.  The factor $2 d_z + d_x +4$ is the width of the unit cell.  Note with the width factor needs $2 (d_z + 1)$ to account for the two patches and $d_x+2$ for the routing space between patches. It suffices to use a distance proportional to $d_x$ (rather than $d_z$) because when merging our lattice surgery patches, the effective distance is maintained.  This leads to a multiplicative routing overhead of
\begin{align}
 O_{(d_z, d_x)} = \frac{2(3 d_x +2)(2 d_z + d_x +4)}{8 d_x d_z} \approx \frac{3}{2}.
 \label{eq:RoutingAsymptot}
\end{align}
where the approximation holds for large asymmetric codes $d_z \gg d_x \gg 4$. \cref{eq:RoutingAsymptot} corresponds to the asymptotic routing overhead since there is extra routing space required shown as green edge padding in \cref{fig:IntroduceIdeas}. However, the cost of edge padding (as a fraction of total costs) vanishes with the number of logical qubits. The routing overhead can be reduced further if not every logical qubit is ready to participate in lattice surgery (e.g. see the core-cache model of Ref.~\cite{CC21}).

\subsection{Performance comparison between twist-based and twist-free lattice surgery}
\label{subsec:PerformanceCompare}

Suppose we wish to measure a multi-qubit Pauli operator $P$ using lattice surgery. The parity of the multi-qubit Pauli measurement is given by a product of stabilizers in the routing space. For instance, for the $P = Y \otimes Y \otimes Y \otimes Y$ measurement in \cref{fig:GaintYY}, the parity of the measurement is given by the product of all $X$-type stabilizers and twist defects in the routing space, in addition to domain wall stabilizers with the $X$-component of the operator incident to the routing space. In what follows, we will refer to timelike failures as a collection of failure mechanisms which can result in the wrong parity outcome of a multi-qubit Pauli measurement implemented via lattice surgery. A logical timelike failure occurs when the incorrect parity of a multi-qubit Pauli is obtained after performing error correction over the full syndrome history of a lattice surgery protocol.

Due to timelike failures in addition to spacelike failures resulting in data qubit errors which anticommute with $P$, the measurement of stabilizers in the routing space needs to be repeated $d_m$ times where $d_m$ is determined from the noise model and size of the computation. In particular, in a Pauli-based computation requiring the measurement of multi-qubit Pauli operators $\{P_1, \cdots, P_{\mu} \}$, in Ref.\cite{CC21} it was shown that $d_m$ can be chosen to satisfy $\mu L a (bp)^{(d_m+1)/2} \le \delta$ where we require each multi-qubit Pauli measurement to fail with probability no greater than $\delta$. Here $L$ is the worst case area of the routing space used for lattice surgery, $\{a,b\}$ are constants and $p$ quantifies the failure probability of all physical operations on the qubits.


 The goal of this section is to quantify the relative performance of twist-based lattice surgery to a lattice surgery protocol free of twist defects. Such a comparison will allow us to quantify the extra algorithm runtime costs associated with twist defects. In particular, the weight-five stabilizers and elongated stabilizers have more fault locations compared to the weight-four stabilizers of the surface code and will thus make timelike failures more likely when measuring multi-qubit Pauli operators containing $Y$ terms. 
In performing the comparison, we use the following biased circuit-level depolarizing noise model
\begin{enumerate}
    \item Each single-qubit gate location is followed by a Pauli $Z$ error with probability $\frac{p}{3}$ and Pauli $X$ and $Y$ errors each with probability $\frac{p}{3\eta}$.
	\item Each two-qubit gate is followed by a $\{ Z\otimes I, I \otimes Z, Z \otimes Z \}$ error with probability $p/15$ each, and a $\{X \otimes I, I \otimes X, X \otimes X, Z \otimes X, Y \otimes I, Y \otimes X, I \otimes Y, Y \otimes Z, X \otimes Z, Z \otimes Y, X \otimes Y, Y \otimes Y \}$ each with probability $\frac{p}{15 \eta}$.
	\item With probability $\frac{2p}{3\eta}$, the preparation of the $\ket{0}$ state is replaced by $\ket{1}=X\ket{0}$. Similarly, with probability $\frac{2p}{3}$, the preparation of the $\ket{+}$ state is replaced by $\ket{-}=Z\ket{+}$.
	\item With probability $\frac{2p}{3\eta}$, a single-qubit $Z$ basis measurement outcome is flipped. With probability $\frac{2p \alpha}{3}$, a single-qubit $X$-basis measurement outcome is flipped.
	\item Lastly, each idle gate location is followed by a Pauli $Z$ with probability $\frac{p}{3}$, and a $\{X,Y\}$ error each with probability $\frac{p}{3\eta}$.
\end{enumerate}
In performing our simulations, we consider two regimes for $X$-basis measurement errors. The first is for $\alpha = 1$ and the second is for $\alpha = 10$. All simulations are performed using $\eta = 100$, so that $Z$ errors are one hundred times more likely than $X$ and $Y$ errors. In what follows, we use a MWPM algorithm \cite{Edmonds65} on graphs with weighted edges to correct errors arising on the logical patches and in the routing space during a lattice surgery protocol. In particular, we use the decoding algorithm described in Section III of Ref.\cite{CC21}.  

\begin{figure}
    \centering
    \includegraphics[width=0.45\textwidth]{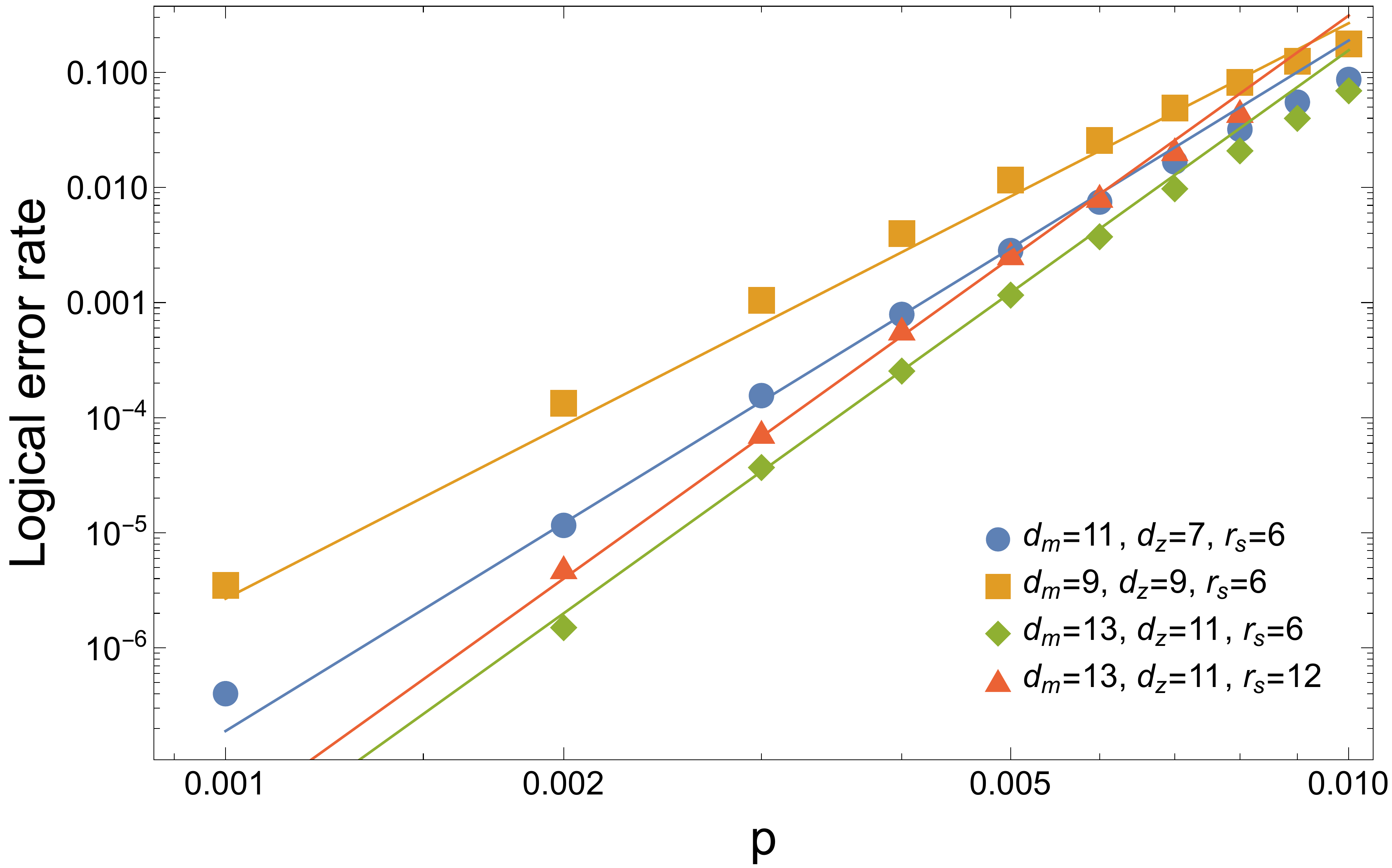}
    \caption{Plot of the timelike logical failure rates of a $Y \otimes Y$ measurement implemented by lattice surgery with twist defects for various values of $d_z$, $d_m$ and $r_s$ with $\alpha = 10$. The results are obtained by preforming $10^7$ Monte Carlo simulations using the noise model described in \cref{subsec:PerformanceCompare}. The solid lines correspond to the best fit polynomial of \cref{eq:10xMeasPoly}. Our fits closely match the data for physical error rates $p \lesssim 7 \times 10^{-3}$. Larger values of $p$ require the inclusion of higher order contributions to the polynomials in \cref{eq:ZZpoly,eq:YYpoly}.  }
    \label{fig:LogicalsDms}
\end{figure}

\begin{figure*}
	\centering
	\subfloat[\label{fig:PlotDelta10x}]{%
		\includegraphics[width=0.5\textwidth]{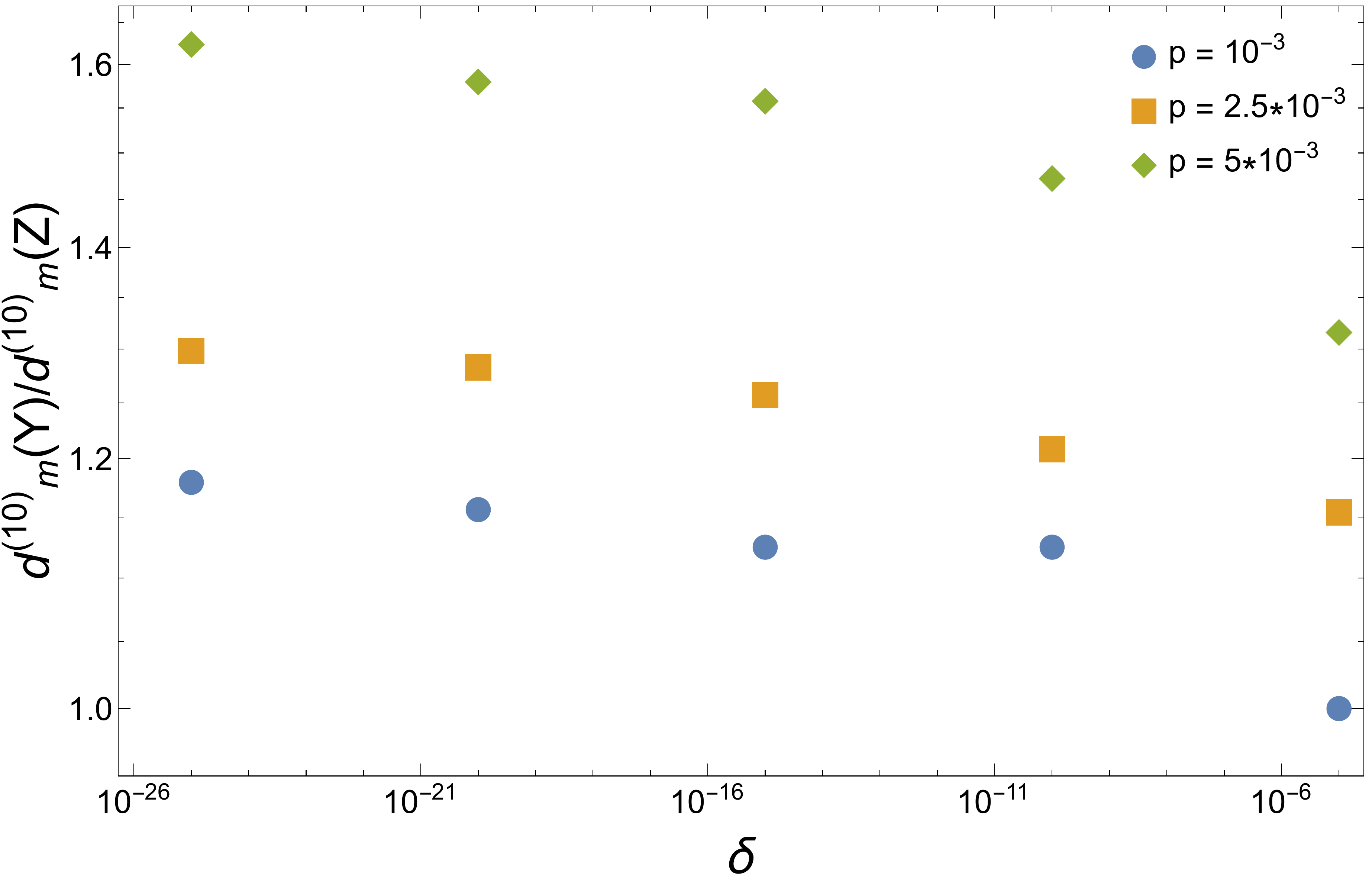}
	}
	\subfloat[\label{fig:PlotDelta1x}]{%
		\includegraphics[width=0.5\textwidth]{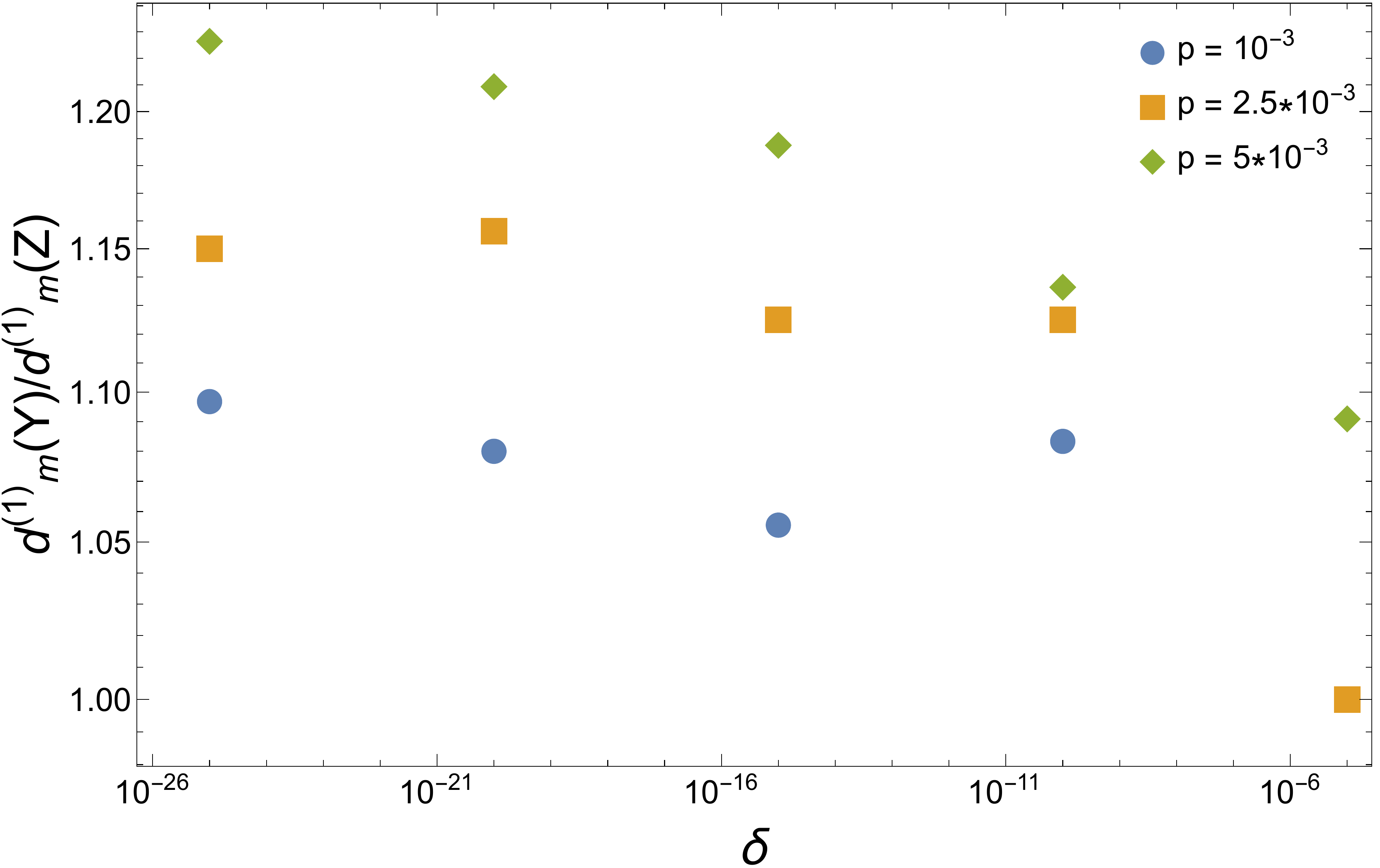}
	}

	\caption{\label{fig:PlotDms} (a) Ratio of $d^{(\alpha)}_m(Y)/d^{(\alpha)}_m(Z)$ for $\alpha = 10$ and various values of the target logical error rate $\delta$ such that both $P^{(YY)}_{L;\alpha}(p) \le \delta$ and $P^{(ZZ)}_{L;\alpha}(p) \le \delta$. The area of the routing space of the $Z \otimes Z$ measurement is taken to be identical to that of the $Y \otimes Y$ measurement. (b) Same as (a) but with $\alpha = 1$. All values were obtained by fixing $r_s = d_x = 7$ and $d_z = 31$. For large $d_z$, the results have very little senstivity to the chosen values of $d_x$, $d_z$ and $r_s$.}
\end{figure*}

For the twist-free simulation of lattice surgery, we consider a $P_1 = Z\otimes Z$ measurement due to the presence of domain walls at the boundaries of the logical patches and routing space. To gain some intuition on the expected scaling of the timelike logical failure rate (i.e. failures which result in the wrong interpretation of the parity of the $P_1$ measurement), suppose that the stabilizers of the merged patch for measuring $Z \otimes Z$ via lattice surgery are measured $d_m$ times. Suppose further that starting in the first syndrome measurement round of the merged patch, a string of $(d_m+1)/2$ consecutive measurement errors occur on one of the stabilizers in the routing space encoding the parity of the $Z \otimes Z$ measurement. Since the series of measurement errors begin in the first round, the inferred parity of the $P_1$ will be incorrect. Further, the minimum-weight path connecting the highlighted vertices\footnote{A vertex $v^{(k)}_j$ of the matching graph used to perform MWPM encodes the parity of the measurement outcome of a given stabilizer $g_j$ in round $k$. The vertex $v^{(k)}_j$ is highlighted if the measurement outcome of $g_j$ changes between consecutive syndrome measurement rounds $k-1$ and $k$.} will not pass through the parity vertices in the first syndrome round thus resulting in a logical timelike failure (see Ref.\cite{CC21} for more details). Additionally, the number of fault locations where timelike failures can occur is proportional to the area of the qubits in the routing space. As such, we propose the following ansatz for the timelike logical failure rate of a $Z \otimes Z$ measurement:
\begin{align}
P^{ZZ}_L(p) = a_1 r_s dz (b_1 p)^{(d_m+1)/2},
\label{eq:ZZpoly}
\end{align}
where $d_z r_s$ is the area of the routing space  between the two domain walls (see \cref{fig:JustYY}), and $\{a_1, b_1 \}$ are constants.

For the measurement of the operator $P_2 = Y \otimes Y$, we expect a similar scaling to the one provided in \cref{eq:ZZpoly}. However, as can be seen in \cref{fig:JustYY,fig:GaintYY}, measuring logical $Y$ operators using twist defects results in narrow vertical strips of length $r_s$ which contain elongated stabilizers. Given the extra fault locations of both twist defects and elongated stabilizers, we consider the following ansatz for timelike logical failures
\begin{align}
P^{YY}_L(p) = P^{ZZ}_L(p) + a_2 r_s (b_2 p)^{(d_m + 1)/2}.
\label{eq:YYpoly}
\end{align}
In \cref{eq:YYpoly}, $P^{ZZ}_L(p)$ is given by \cref{eq:ZZpoly} with $r_s dz$ replaced by the area of the routing space used for the $Y \otimes Y$ measurement. In other words, the additional ``cost" of performing twist-based lattice surgery is an additive error contribution proportional to the height $r_s$ containing twist defects and elongated stabilizers. For a multi-qubit Pauli measurement $P = P_1 \otimes P_2 \otimes \cdots \otimes P_k$ containing $t \le k$ $Y$ terms, we expect the timelike failure probability to scale as 
\begin{align}
P^{P}_L(p) = P^{\tilde{P}}_L(p) + t a_2 r_s (b_2 p)^{(d_m + 1)/2},
\label{eq:ansatzMultiY}
\end{align}
for some constants $\{ a_2, b_2 \}$ where $\tilde{P} = Z_1 \otimes Z_2 \otimes \cdots \otimes Z_k$. In \cref{eq:ansatzMultiY}, we assume that the merged surface code patch for the $P$ measurement has the same area as the merged surface code patch for the $\tilde{P}$ measurement. 

Performing full circuit-level simulations using the biased noise model and decoding algorithm described above, we obtained the following best fit polynomials describing the timelike logical failure rates using the ansatzes of \cref{eq:ZZpoly,eq:YYpoly}
\begin{align}
P^{YY}_{L;\alpha=10}(p) =& 0.0365 r_s (d_x+d_z)(49.44p)^{(d_m+1)/2} \nonumber \\
&+0.0514 r_s(88.145p)^{(d_m+1)/2},
\label{eq:10xMeasPoly}
\end{align}
and
\begin{align}
P^{YY}_{L;\alpha=1}(p) =& 0.0162 r_s (d_x+d_z)(22.43p)^{(d_m+1)/2} \nonumber \\ 
&+0.0280 r_s(36.345p)^{(d_m+1)/2}.
\label{eq:1xMeasPoly}
\end{align}
The first term in \cref{eq:10xMeasPoly,eq:1xMeasPoly} can be used to extract the best fit polynomials for $Z \otimes Z$ lattice surgery with a routing space area given by $r_s(d_x+d_z)$. In \cref{fig:LogicalsDms}, we show how the polynomial in \cref{eq:10xMeasPoly} fits the data obtained from our Monte Carlo simulations for various values of $d_z$, $d_m$ and $r_s$ (with fixed $d_x=7$).  

Now suppose that we wish to implement a quantum algorithm so that all lattice surgery operations fail with probability no more than $\delta$. Hence we must have that $P^{(P)}_L \le \delta$ which will fix the required value of $d_m$ for each multi-qubit Pauli measurement. Using the results of \cref{eq:10xMeasPoly,eq:1xMeasPoly}, in \cref{fig:PlotDms}, we plot the ratios $d^{(\alpha)}_m(Y)/d_m^{(\alpha)}(Z)$ for $\alpha = 10$ and $\alpha = 1$. Here $d^{(\alpha)}_m(Y)$ corresponds to the minimum value of $d_m$ such that $P^{(YY)}_{L;\alpha}(p) \le \delta$. Note that in computing $d^{(\alpha)}_m(Y)$, the extension of the logical patch to obtain a horizontal $Y$ boundary is done simultaneously with the gauge fixing step as explained in \cref{sec:FaultToleranceGauge}. As such, $d^{(\alpha)}_m(Y)$ counts all syndrome measurement rounds from the gauge fixing step until the round when the patches are split. Similarly, $d^{(\alpha)}_m(Z)$ corresponds to the minimum value of $d_m$ such that $P^{(ZZ)}_{L;\alpha}(p) \le \delta$ for a lattice surgery patch where the routing space has the same area as the one used for a $Y \otimes Y$ measurement. Hence, the ratio $d^{(\alpha)}_m(Y)/d_m^{(\alpha)}(Z)$ allows us to quantify the extra multiplicative time cost of measuring $Y \otimes Y$ via lattice surgery using twist defects compared to a lattice surgery protocol for measuring $Z \otimes Z$ which does not require the use of twist defects. A few remarks are in order.

First, the ratio $d^{(\alpha)}_m(Y)/d_m^{(\alpha)}(Z)$ is highly insensitive to the chosen values of $d_x$, $r_s$ and $d_z$. As such, the plots in \cref{fig:PlotDms} were obtained by fixing $d_z = 31$ and $d_x=r_s=7$. We chose the value $d_x = 7$ since for the biased noise model considered here, it was shown in Ref.\cite{CC21} that $d_x=7$ was sufficient to implement the Hubbard model \cite{HubbardOrig,CampbellHubbard} for lattice sizes $L \le 32$.

Second, comparing the results in \cref{fig:PlotDelta10x} to those in \cref{fig:PlotDelta1x}, it can be seen that higher measurement failure rates increase the runtime for performing lattice surgery with twist defects. Further, for physical noise rates $p \approx 10^{-3}$, the ratio $d^{(\alpha)}_m(Y)/d_m^{(\alpha)}(Z) \approx 1.2$ when $\alpha = 10$. However, $d^{(\alpha)}_m(Y)/d_m^{(\alpha)}(Z)$ increases fairly quickly as $p$ approaches the $10^{-2}$ regime which is near the threshold for logical timelike failures.

It can be seen in \cref{fig:PlotDelta1x} that the ratio $d^{(1)}_m(Y) / d^{(1)}_m(Z)$ for $p = 10^{-3}$ increases when going from $\delta = 10^{-15}$ to $\delta = 10^{-10}$. The absolute values of $d^{(1)}_m(Y)$ and $d^{(1)}_m(Z)$ are larger when $\delta = 10^{-15}$. However, for both values of $\delta$, we only require $d^{(1)}_m(Y) = d^{(1)}_m(Z) + 1$ to achieve the desired target logical failure rates. As such, the fraction $d^{(1)}_m(Y) / d^{(1)}_m(Z)$ is larger when the absolute values of $d^{(1)}_m(Y)$ and $d^{(1)}_m(Z)$ decrease.

Lastly, we remark that for multi-qubit Pauli measurements requiring more than two $Y$ terms, the ratios $d^{(\alpha)}_m(Y) / d^{(\alpha)}_m(Z)$ obtained in \cref{fig:PlotDms} would only increase in a meaningful way for very large values of $t$ when using the result of \cref{eq:ansatzMultiY} due to the exponential scaling of the second term. As such, for the noise rates considered in \cref{fig:PlotDms}, there could be rare cases where a multi-qubit Pauli measurement containing a very large number of $Y$ terms would benefit from a twist-free approach such as the one considered in \cite{CC21}. 

\section{Conclusion}

In Ref.\cite{CC21}, the twist-free scheme for measuring multi-qubit Pauli operators containing $Y$ terms results in an approximately $2 \times$ slowdown in algorithm runtime due to the need to sequentially measure multi-qubit Pauli operators containing pure $X$ terms, followed by a multi-qubit Pauli measurement containing pure $Z$ terms. Our results thus suggest that, assuming a hardware architecture satisfying the connectivity constraints described in \cref{fig:Connectivity}, twist-based lattice surgery will outperform the twist-free version for physical noise rates $p  \lesssim 5 \cdot 10^{-3}$. However, for noise rates close to the threshold value $\sim 10^{-2}$, we expect twist-free based lattice surgery to achieve lower algorithm runtime relative to lattice surgery protocols using twist defects to measure $Y$ operators. An open question is whether high threshold surface code implementations are possible using homogeneous architectures and less connectivity as was assumed in \cref{fig:Connectivity}.

The above conclusions also depend on the noise model. For instance, comparing \cref{fig:PlotDelta10x,fig:PlotDelta1x}, it is clear that higher measurement failure rates results in longer $Y \otimes Y$ measurements compared to $Z \otimes Z$  measurements with an identical area. As such, for noise models where measurement and two-qubit gate failures are very high relative to other failure mechanisms, runtimes of twist-free lattice surgery protocols could potentially be smaller compared twist-based lattice surgery protocols.

\appendix

\section{Stabilizer measurement circuits for twist defects in the bulk}
\label{appendix:StabMeasTwist}

\begin{figure}
	\centering
	\subfloat[\label{fig:StabParity}]{%
		\includegraphics[width=0.3\textwidth]{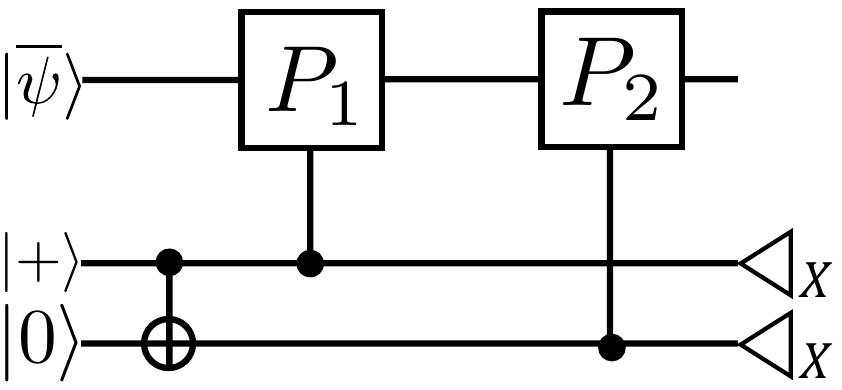}
	}
	\vfill
	\subfloat[\label{fig:StabCorrP}]{%
		\includegraphics[width=0.3\textwidth]{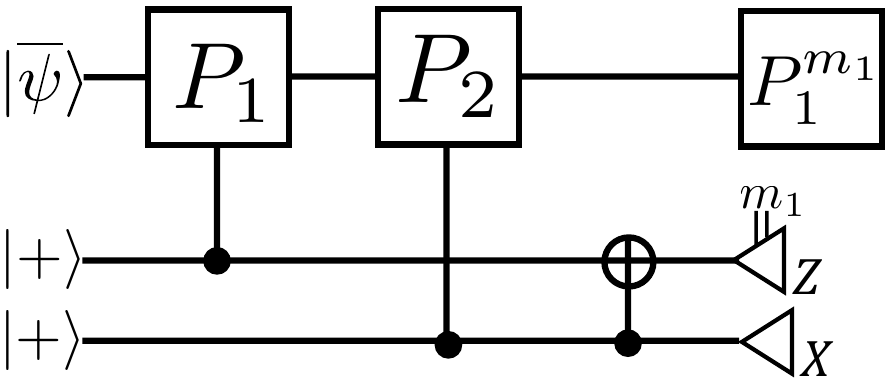}
	}

	\caption{\label{fig:StabMeasCircuitsTwist} (a) Circuit used to measure the weight-five stabilizers of the twist defects located on the left vertical strip of \cref{fig:GaintYY}. A GHZ state is prepared, and the parity of the stabilizer measurement outcome is given by the product of the two $X$-basis measurement outcomes. (b) Second type of circuit used to measure the weight-five stabilizers of the twist defects located on the right vertical strip in \cref{fig:GaintYY}. In this case, the $X$-basis measurement outcome gives the parity of the stabilizer measurement. A correction $P_1^{m_1}$ (or equivalently $P_2^{m_1}$) is applied based on the $m_1$ $Z$-basis measurement outcome. For both circuits, $P_1P_2$ is the stabilizer being measured.}
\end{figure}

In this appendix, we describe the two types of circuits used to measure elongated checks and the weight-five stabilizers corresponding to twist defects on both the left and right vertical strips in the routing space of \cref{fig:GaintYY}. We presented one such example in \cref{fig:Connectivity} and here give a more general procedure. Such circuits are shown in \cref{fig:StabParity,fig:StabCorrP}, where $\ket{\overline{\psi}}$ is the encoded state protecting the data during lattice surgery. $P_1$ and $P_2$ are Pauli operators with $P_1P_2$ corresponding to the stabilizer being measured. 

\begin{figure*}
    \centering
    \includegraphics[width=\textwidth]{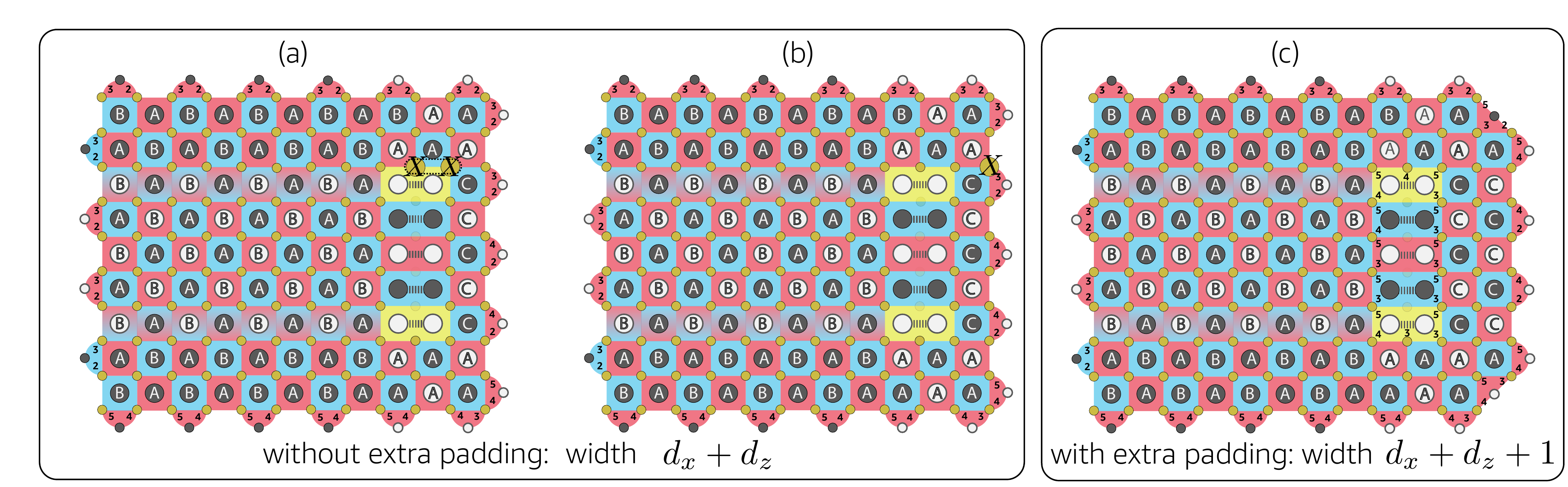}
    \caption{ In (a) and (b), we show two identical surface code patches obtained after performing a $Y \otimes Y$ measurement by lattice surgery. The width of both patches is $d_x + d_z$. (a) We consider a failure arising at the controlled-$Y$ gate for measuring the top twist defect during the first syndrome measurement round of the merged patch. The failure results in an $X\otimes X$ error on two data qubits as shown in the figure. Since the measurement of the twist defect is random during the first round of the merge, and no changes are detected in subsequent rounds (assuming no other faults), only the vertex corresponding to the blue $Z$-type stabilizer which anti-commutes with the error is highlighted. (b) A weight-one $X$ error occurs on the data qubit shown in the figure resulting in the same syndrome measurement outcome as the one obtained in (a). Since both types of failures have the same syndrome measurement outcome and are parallel to a logical $X$ operator of the surface code patch, a single failure on this lattice can lead to a logical fault. (c) Patch for measuring $Y \otimes Y$ with width $d_x + d_z + 1$. The patch uses extra padding to allow both leading order failures considered in this example to have distinct syndromes. Such a setting thus requires the logical patches to have width $d_x+1$ prior to performing lattice surgery.}
    \label{fig:ExampleError}
\end{figure*}

Without loss of generality, assume the data qubits is in an eigenstate of $P_1P_2$ so that $P_1P_2\ket{\overline{\psi}}=\pm \ket{\overline{\psi}}$.
For the circuit in \cref{fig:StabParity}, it is straightforward to show that the state prior to the measurements is given by $\ket{\psi}_f =  (\ket{0,0}  \pm \ket{1,1})/\sqrt{2} \ket{\overline{\psi}}$.  The ancilla state $(\ket{0,0} \pm \ket{1,1})/\sqrt{2}$ is a eigenstate of $X \otimes X$ with eigenvalue $\pm 1$. As such, the product of the two $X$-basis measurement outcomes of the ancilla qubits gives the stabilizer measurement outcome.

For the circuit in \cref{fig:StabCorrP}, the state prior to the measurement is given by 
\begin{align}
 \ket{\psi}_f = (\ket{0,0} + \ket{0,1}P_2 +\ket{1,0}P_1 + \ket{1,1}P_1 P_2) \ket{\overline{\psi}} 
\end{align}
which we can express as 
\begin{align} \nonumber
  \ket{\psi}_f & = \ket{0}(\ket{0}+\ket{1}P_1P_2)\ket{\overline{\psi}} + \ket{1}(\ket{0}P_1+\ket{1}P_2)\ket{\overline{\psi}} \\  \nonumber
& = \ket{0}(\ket{0}+\ket{1}P_1P_2)\ket{\overline{\psi}} +P_1 \ket{1}(\ket{0}+\ket{1}P_2)\ket{\overline{\psi}} \\ 
& = \ket{0}\ket{\pm}\ket{\overline{\psi}} + P_1\ket{1}\ket{\pm}\ket{\overline{\psi}}  
\end{align}
Therefore, measuring the second ancilla qubit in the $X$ basis will reveal the desired $\pm 1$ eigenvalue.  However, we also have a $P_1$ Pauli correction required depending on the value of the first ancilla qubit measured the $Z$ basis.

Lastly, we point out that using flag based methods \cite{CR1,CR2,CB18,TCL20,ReichardtFlag,ChamberlandMagic,chamberland2020very,ChamberlandHeavyHex,ChamberlandColorCode,CR3,TD21Ed,PR21Flag,ExperimentalFlagDem,TRColor21} to measure the elongated checks and weight-five twist defects would require an extra two-qubit gate between the ancillas instead of a single CNOT. As such, an extra idling time would be required resulting in slower algorithm runtimes. 

\section{Minimum distance requirement of twist-based lattice surgery}
\label{app:Subtleties}

In this appendix, we show that twist-based lattice surgery protocols require extra padding to correct arbitrary $X$ errors arising from $(d_x - 1)/2$ failures. 

An example of a weight-one $X$ error being indistinguishable from a weight-two $X$ error (both arising from a single failure with probabilities proportional to $p$) for a lattice surgery protocol measuring $Y \otimes Y$ is shown in \cref{fig:ExampleError}. The lattices in \cref{fig:ExampleError}(a) and (b) have width $d_x + d_z$ where $d_x = 3$ and $d_z = 9$. The lattice in \cref{fig:ExampleError}(c) has extra padding making its width $d_x + d_z + 1$. For the figure containing the lattices of width $d_x + d_z$, consider first the lattice in \cref{fig:ExampleError}(a). A weight-two $X$ data qubit error occurs during the first syndrome measurement round of the merged patch (i.e. the first syndrome measurement round when gauge fixing) due to a controlled-$Y$ gate failure used for measuring the top twist defect. The error anticommutes with the weight-five yellow plaquette in addition to a blue weight-four $Z$-type stabilizer. However, the measurement outcome of the weight-five operator is random in the first round of the merge, and no change in measurement outcomes are observed in subsequent syndrome measurement rounds. As such, only the vertex associated with the weight-four $Z$-type stabilizer in the first round of the merge is highlighted. The observed syndrome measurement outcome is identical to the one that would be obtained if a weight-one $X$ error, as shown in \cref{fig:ExampleError}(b), occurred during the first round of the merge. Since both a weight-one and weight-two $X$ errors result in the same syndrome measurement history, and $d_x = 3$, this example shows that a single failure occurring during the first syndrome measurement round when merging the surface code patches to measure $Y \otimes Y$ can lead to a logical failure. However, by increasing the $d_x$ distance by one using extra padding, as shown in \cref{fig:ExampleError}(c), such failure mechanisms can be distinguished. 

More generally, suppose we perform lattice surgery using twist defects to measure $Y$ operators and the initial logical patches prior to the merge have odd distance $d_x$. In such settings, fewer than $(d_x + 1)/2$ failures occurring during the first syndrome measurement round of the merged patch can result in a logical $X$ error. To ensure that up to $(d_x + 1)/2$ failures never result in a logical failure, the logical patches can always be padded with an extra column of stabilizers thus increasing their $d_x$ distance by one. Depending on the noise bias and size of the quantum algorithm, the extra padding might not be necessary. For instance, the simulations performed in \cref{subsec:PerformanceCompare} with a noise bias of $\eta = 100$ used the odd distance $d_x = 7$ which is sufficient for implementing the Hubbard model with lattice sizes $L \le 32$. Further, even though for odd distances $d_x$ the logical $X$ failure probability scales as $\binom{N}{(d_x - 1)/2} p^{(d_x - 1)/2}$, the combinatorial coefficient $\binom{N}{(d_x - 1)/2}$ is very small relative to the combinatorial coefficient of higher order failure mechanisms since only $(d_x - 1)/2$ failures in the first round of the merged patch at very specialized regions near the twist defects can lead to a logical failure. 

\begin{figure*}[t!]
    \centering
    \includegraphics[width=\textwidth]{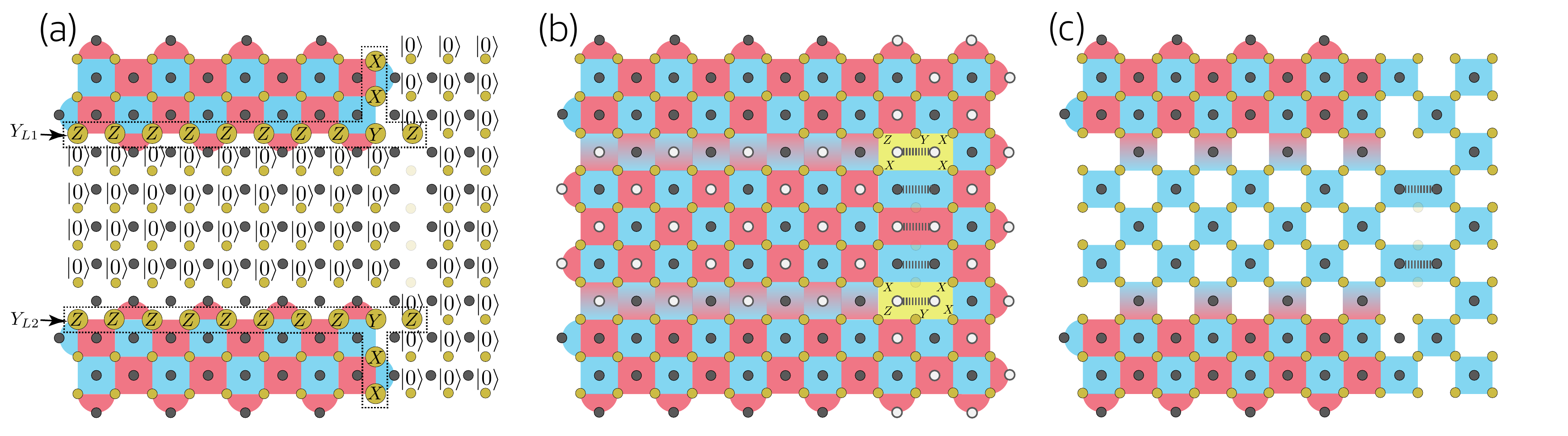}
    \caption{Illustration showing an efficient protocol for measuring $Y \otimes Y$ fault-tolerantly. In (a), we illustrate the stabilizers for the initial configuration of the logical patches including single qubit $Z$ for $\ket{0}$ states.  In (b), we show the stabilizers of the merged patched obtained via lattice surgery. In (c), we show the stabilizers of the sub-system code that (a) and (b) can be regarded as a gauge fixing of.
    }
    \label{fig:FaultToleranceProof}
\end{figure*}

\section{Fault tolerance proof}
\label{sec:FaultToleranceGauge}

Given a rectangular logical patch, in Ref.~\cite{litinski2019game} Litinski showed that the Pauli $Y$ operator can be measured in several steps. We start with two separate surface code patches, the patch to be measured and a $\ket{0_L}$ ancilla. The next step is to extend the patch by moving the corner to express a minimum weight representative of the logical $Y$ operator along a straight line. Afterwords, a twist defect is used to measure $Z \otimes Y$ with an ancilla prepared in a logical $\ket{0}$ state. The last step implements destructive measurements on qubits in the ancilla region in order to return to the original code space. Litinski argued that the first and second steps require $d$ rounds of error correction, and all other steps require effectively zero time.  We will refer to the second step as the \textit{extension} step. If the extension step is required for a single $Y$ logical measurement, it would seem natural to prefer this for $Y \otimes Y$ and more general logical measurements.  However, the extension step effectively doubles the execution time for lattice surgery and also demands additional routing space. 

Here we argue that the extension step can be skipped while retaining all the fault-tolerance properties of lattice surgery at the phenomenological level.  We borrow the gauge-fixing ideas of Ref.~\cite{Vuillot_2019} and apply them to the example of a $Y \otimes Y$ measurement as sketched in \cref{fig:FaultToleranceProof}.  

The initial state of \cref{fig:FaultToleranceProof}(a) has some set of stabilizers $\mathcal{S}_{\mathrm{old}}$ that includes all the stabilizer generators for the two logical patches. Note the data qubits in the routing space prepared in the $\ket{0}$ state. Unconventionally, but crucial to our proof, we regard the $Z$ Pauli for each $\ket{0}$ qubit as being part of the stabilizer $\mathcal{S}_{\mathrm{old}}$.  This further allows us to choose an unusual representative for the $Y_{L1}$ and $Y_{L2}$ logical operators (each differs by a $Z$ from $\mathcal{S}_{\mathrm{old}}$ relative to the standard representatives for $Y_{L1}$ and $Y_{L2}$).

The merged state of \cref{fig:FaultToleranceProof}(b) illustrates the stabilizers $\mathcal{S}_{\mathrm{new}}$  during the merge step (having skipped any extension step).  Multiplying the stabilizers with white vertices gives our representative for $Y_{L1} \otimes Y_{L2}$ from the initial configuration and therefore these lattice surgery operations do measure the desired logical operator.  If we had not included the single-qubit $Z$ operators in $\mathcal{S}_{\mathrm{old}}$, we would not have measured $Y_{L1} \otimes Y_{L2}$, so this is the key formal trick to prove fault-tolerance.

Lastly, Ref.~\cite{Vuillot_2019}  prescribes that we consider the sub-system code with stabilizer $\tilde{\mathcal{S}}= \mathcal{S}_{\mathrm{old}} \cap \mathcal{S}_{\mathrm{new}}$ and gauge group $\tilde{\mathcal{G}}= \langle \mathcal{S}_{\mathrm{old}} , \mathcal{S}_{\mathrm{new}} \rangle$.  The protocol is then (phenomenologically) fault-tolerant provided that this sub-system code retains the distance of the initial code.  We illustrate the local generators for $\tilde{\mathcal{S}}$ in \cref{fig:FaultToleranceProof}(c). The logical operators for the sub-system code must commute with the stabilizer group and not be contained in the gauge group. Examining \cref{fig:FaultToleranceProof}(c) one finds that the code distance is indeed maintained at $d_z$ for pure $Z$ errors and at least $d_x$ for all other types of errors.

\bibliography{TwistLatticeSurgery}

\end{document}